\documentclass[aps,prc,showpacs,amsmath,amssymb,superscriptaddress]{revtex4-1}

\usepackage{graphicx}
\usepackage{amsmath}
\usepackage{hyperref}
\usepackage{dcolumn}
\usepackage{bm}
\usepackage{color}
\usepackage{nicefrac}

\begin{document}

\title{Ising-QCD phenomenology close to the critical point}

\author{N.~G.~Antoniou}
 \email[]{nantonio@phys.uoa.gr}
\affiliation{Faculty of Physics, University of Athens, GR-15784 Athens, Greece}

\author{F.~K.~Diakonos}
 \email[]{fdiakono@phys.uoa.gr}
\affiliation{Faculty of Physics, University of Athens, GR-15784 Athens, Greece}

\date{\today}

\begin{abstract}
\noindent
We employ the recently introduced Ising-QCD partition function (N.~G. Antoniou {\it et al.}, Phys. Rev. D 97, 034015 (2018)) to explore in detail the behaviour of the moments of the baryon-number, within the critical region around the critical endpoint. Our analysis is based on the relation of finite-size scaling in real space with intermittency in transverse momentum space. It demonstrates in practice the recent observation (N.~G. Antoniou {\it et al.},  Phys. Rev. D 97, 034015 (2018)) that combined measurements of the intermittency index $\phi_2$ and the freeze-out parameters $\mu_b$ (baryochemical potential), $T$ (temperature), provide us with a powerful tool to detect the critical point. We also show that the finite-size scaling (FSS) region, as a part of the critical region, is very narrow in both the chemical potential and the temperature direction, even for light nuclei. Furthermore, using published experimental results for $(\mu_b,T,\phi_2)$ in A+A collisions at $\sqrt{s_{NN}}=17.2$ GeV (NA49 experiment, CERN-SPS), we are able to make a set of predictions for the freeze-out states of Ar + Sc and Xe + La collisions at the same energy in the NA61/SHINE experiment (CERN-SPS). In particular, we find that the Ar + Sc system freezes out outside the FSS region but very close to its boundary, a property which may leave characteristic traces in intermittency analysis.

\end{abstract}

\pacs{} 

\maketitle
\section{Introduction}

The hunt for the QCD critical endpoint (CEP), remnant of the chiral symmetry breaking, in modern ion collision experiments approaches its apex. During the last decade it became evident that in the phase diagram of strongly interacting matter of finite size, as it is the case for colliding nuclei, the CEP is replaced by an extended critical region in the temperature-baryon chemical potential plane \cite{Gavai2016}. Recent experimental measurements of the baryon number non-Gaussian kurtosis $\kappa$ (times variance $\sigma^2$) in the Beam-Energy Scan (BES-I) program at BNL-RHIC \cite{Luo2017} display a broad minimum in the range $15$ GeV $< \sqrt{s} <$ $40$ GeV as a function of the beam energy. This non-monotonic behaviour is interpreted as a potential trace of criticality \cite{Stephanov2011,Luo2017}. Two years ago, measurements of the intermittency index $\phi_2$, employing proton transverse momenta in A+A collisions at beam energy $\sqrt{s}=17.2$ GeV in NA49 experiment (CERN-SPS), provided an indication for the occurrence of critical fluctuations in the Si + "Si" freeze-out state \cite{Anticic2015}. These two, seemingly incompatible results if the critical region is not unnaturally wide, may constitute an invaluable guide for locating the CEP, as we will show in the analysis presented in this paper. 

Our aim is to use the recently introduced Ising-QCD partition function \cite{Antoniou2017} to investigate in some detail the phenomenology of the critical region in terms of finite-size scaling (FSS) in baryon number moments and its relation to intermittency in momentum space. In our treatment we consider the freeze-out states generated in relativistic ion collision experiments in thermal and chemical equilibrium as supported also by the experimental findings. Therefore, we expect that dynamical effects related to the relaxation process and studied recently in the literature \cite{Herold2014,Stephanov2017}, are not necessarily the guiding rules in the phenomenology of the baryonic fluid in the critical region, which is governed by equilibrated critical fluctuations.
     
In particular we will demonstrate that the FSS region, which is a subregion of the critical one, is very narrow along the temperature direction, particularly for large size ions. This suggests that the formation of freeze-out states within the scaling region of the critical point, is visible in central collisions of relatively light ions ($A \approx 27 \pm 4$) with colliding energy $\sqrt{s}\approx 17.2~\mathrm{GeV}$. We discuss the impact of our analysis on expected results for the most important running ion experiments: the NA61/SHINE (SPS, CERN) and the STAR (BNL, RHIC). We find that accurate measurements of the intermittency index $\phi_2$ in Ar + Sc central and peripheral collisions at highest colliding energy ($\sqrt{s}=17$ GeV) of the running NA61/SHINE experiment (CERN-SPS) could provide helpful information for the location of the QCD CEP. On the opposite side, due to the narrowness of the critical region along the baryochemical direction \cite{Antoniou2017}, the beam energy scan in BES program at RHIC (BNL) cannot resolve the critical region. Such a task would require much denser covering of the colliding energy range than that available in BES I. Our analysis reveals that in general the strategy of using only ions of large size is not optimal for the CEP search. In addition the energy scan is also inadequate since the scaling region around the CEP is very narrow. In fact, the size scan becomes decisive.

Our paper is organized as follows: in section 2 we briefly present the theoretical framework (Ising-QCD partition function) used in our subsequent phenomenological analysis. More extensive presentation can be found in \cite{Antoniou2017}. In section 3 we present the connection of finite-size scaling in configuration space with intermittency in transverse momentum space emphasizing on the constraints imposed to the temperature for freezing out within the FSS region. In section 4 we perform a detailed phenomenological analysis, combining existing experimental results with the theoretical framework developed in \cite{Antoniou2017}, in order to obtain a set for predictions concerning measurable quantities, related to critical fluctuations, in the NA61/SHINE and STAR experiments. Finally, in section 5 we present our conclusions and give also some perspectives in forthcoming experimental searches for the CEP.  

\section{Ising-QCD critical thermodynamics}

We start our study recalling the basic thermodynamics of the critical fluid, which signals the QCD CEP, adopting the viewpoint of \cite{Antoniou2017}. Due to baryon number conservation, the slow mode of the order parameter, dominating at macroscopic scales, is the baryon number density $n_b$. Since the CEP is expected to belong to the 3-d Ising universality class, as supported by several theoretical works\cite{Gavin1994,Stephanov1998,Halasz1998,Berges1999,Karsch2001}, the  fluctuations of $n_b$ near the critical point are described by the universal effective action found by Monte-Carlo simulations in \cite{Tsypin1994}. Thus, the thermodynamics of the strong interacting matter close to the CEP are determined by the Ising-QCD partition function:  
\begin{equation}
\mathcal{Z}=\displaystyle{\sum_{N=0}^M} \zeta^N \exp\left[-\frac{1}{2} \hat{m}^2 \frac{N^2}{M}-g_4 \hat{m} \frac{N^4}{M^3} - g_6 \frac{N^6}{M^5}\right]
\label{eq:1}
\end{equation}
where $N$ is the baryon number in volume $V=M \beta_c^3$, $\beta_c=1/k_B T_c$ being the characteristic length scale of the system fixed by the critical temperature $T_c$. The dimensionless variable $M$ quantifies the size of the considered baryonic system. The dimensionless universal couplings $g_4=0.97 \pm 0.02$, $g_6=2.05 \pm 0.15$ are calculated in \cite{Tsypin1994} while $\hat{m}=\beta_c m$ is related to the correlation length $\xi=m^{-1}$ of the infinite system. Finally, $\zeta=\exp[\frac{\mu_b-\mu_c}{k_B T_c}]$ is the fugacity and ($\mu_c$) $\mu_b$ is the (critical) baryochemical potential. The partition function (\ref{eq:1}) provides a valid description of the critical fluid thermodynamics within the critical region where a distinction between hadronic and quark phase is not possible. As $T$ approaches $T_c$ the fluid enters into the finite-size scaling (FSS) region for which $\xi > V^{1/d}$ (with $d$ the topological dimension, here $d=3$). Defining $t_{\pm} = \vert \frac{T_{\pm} - T_c}{T_c} \vert$, the border of the FSS region in the temperature axis is given by:
\begin{equation}
t_{\pm}=(\frac{\xi_{0,\pm}}{\beta_c})^{1/\nu} M^{-1/d \nu}
\label{eq:2}
\end{equation}
with $\nu=\frac{2}{3}$ the correlation length critical exponent for the 3-d Ising. The index "$\pm$" is used to indicate that there are different correlation length amplitudes above ($\xi_{0,+}$) and below ($\xi_{0,-}$) the critical temperature $T_c$ with a universal ratio $\frac{\xi_{0,+}}{\xi_{0,-}}\approx 2$ in the 3-d Ising class \cite{Huang}. 
As a consequence, the FSS region for $T > T_c$ is more extended in the temperature direction than the corresponding region for $T < T_c$. A quantitative estimate of the width of the FSS region requires the knowledge of the critical temperature $T_c$ and one of the amplitudes $\xi_{0,+}$ or $\xi_{0,-}$. 

\section{Finite-size scaling and Intermittency in momentum space}

According to the Ising-QCD partition function (\ref{eq:1}), within the FSS region in the $(mu_b,T)$-plane, the integrated baryon number density moments, as shown in \cite{Antoniou2017}, very close to CEP obey the scaling relation(s):
\begin{equation}    
\langle N^k \rangle \sim M^{k q}~~~~;~~~~q=d_F/d,~k=1,2,..
\label{eq:3}
\end{equation}
with $d$ the embedding (topological) dimension and $d_F$ the fractal dimension related to the baryon number density critical fluctuations.
The latter, as discussed in earlier work \cite{Antoniou2006}, imply also a local scaling behaviour of the baryon number density-density correlation function:
\begin{equation}
\langle n_b(\mathbf{r}) n_b(\mathbf{r_0}) \rangle \sim \vert \mathbf{r} - \mathbf{r_0} \vert^{-d(1-q)}
\label{eq:4}
\end{equation}
Within the FSS region, the scaling law (\ref{eq:4}) applies also for large $\vert \mathbf{r} - \mathbf{r_0} \vert$, i.e. at scales of the order of the system's size. This property has an important consequence: the power-law behaviour of eq.~(\ref{eq:4}) is transferred to an analogous behaviour in transverse momentum space for small momenta differences through a Fourier transform \cite{Antoniou2016}, as dictated by critical opalescence. Adapted to the geometry of the fireball created in relativistic ion collisions, this line of thoughts leads to the appearance of a power-law behaviour of the baryon number density-density correlation function in transverse momentum space \cite{Antoniou2006,Antoniou2016,Antoniou2017} of the form:
\begin{equation}
\langle n_b(\mathbf{k}) n_b(\mathbf{k_0}) \rangle \sim \vert \mathbf{k} - \mathbf{k_0} \vert^{-\frac{2}{3}dq}
\label{eq:5}
\end{equation}  
which is directly observable through intermittency analysis. The latter requires the calculation of the second factorial moment of the proton number density within $M_{\perp}$ small cells of transverse momentum space and looking for the scaling behaviour:
\begin{equation}
F_2(M_{\perp}) \sim \left(M_{\perp}\right)^{\phi_2}~~~~;~~~~\phi_2=\frac{1}{3}qd~,~q=\frac{5}{6}~,~d=3
\label{eq:6}
\end{equation} 
in terms of the number of cells $M_{\perp}$ \cite{Antoniou2006}. Thus, the exponent $q$ of FSS in real space is directly linked to the intermittency index $\phi_2$ in transverse momentum space: $\phi_2=q$ and therefore it is an observable quantity characterizing the scaling region. In fact, as discussed in \cite{Antoniou2017}, this link between scaling at large distances in configuration space and intermittency (scaling of factorial moments) in transverse momentum space, holds for the entire FSS region, despite that the value of $q$ varies, as we depart from the immediate neighborhood of the CEP, becoming $\tilde{q} \neq d_F/d$. Of course in this more general case: $\phi_2=\tilde{q}$. As argued in \cite{Antoniou2017} the critical region associated with the QCD CEP is determined by the condition:
\begin{equation}
\frac{3}{4} < \tilde{q} < 1
\label{eq:7}
\end{equation}
while the FSS region is the sub-domain of the critical region for which both eq.~(\ref{eq:7}), as well as the condition that the correlation length of the infinite system is greater than the characteristic system's size, are valid. In the context of intermittency in transverse momentum space we are interested in the transverse size of the system. In the case of colliding ions, the transverse size of the formed fireball can be at best estimated in the central rapidity region where transverse and longitudinal coordinates decouple \cite{Antoniou2006}. Then, a natural measure for the average transverse size is obtained by  $R_{\perp}=\sqrt{S}$ with $S$ the average transverse area of the formed fireball. Assuming radial symmetry, the latter can be estimated as: $S=\pi R^2$ with $R=R_0 A^{1/3}$ and $R_0 \approx 1.25~\mathrm{fm}$. To calculate the infinite system's correlation length $\xi$ and compare with $R_{\perp}$ we need to know the critical temperature $T_c$ and the non-universal amplitude $\xi_{0,+}$, as also discussed in the previous section.    

To determine $T_c$ we employ recent results relating the actual critical temperature at finite chemical potential to the (pseudo)critical temperature of Lattice QCD at zero chemical potential by a a factor $0.94 \pm 0.01$ \cite{Gavai2016}. This information, combined with latest accurate results on Polyakov loop and strange quark susceptibilities at zero chemical potential \cite{Datta2017}, restricts the value of the critical endpoint temperature in the range $160~\mathrm{MeV} < T_c < 165~\mathrm{MeV}$. Here we will use the value $T_c=163~\mathrm{MeV}$ as a representative one. Furthermore, for determining the amplitude $\xi_{0,+}$ we assume that the correlation length at $T=2 T_c$ is significantly smaller than the proton radius $1.25~\mathrm{fm}$, using indicatively the value $\xi(2 T_c)=\xi_{0,+} \approx \frac{1}{2}~\mathrm{fm}$. With this choice we can find, for different nuclear systems, the temperatures $T_{\pm}$ which determine the upper $(+)$ and lower $(-)$ limits of the FSS region along the temperature axis, compatible with the Ising-QCD partition function (\ref{eq:1}). The results are summarized in Table I, where the transverse size of the corresponding nuclei (in fm) is also given:
\begin{table}
\begin{tabular}{| l | c | c | c |}
\hline 
System (A) & $T_+$ (MeV) & $T_-$ (MeV) & $R_{\perp}$ (fm)\\
\hline
\hline
Be   (8) & 169.2 & 160.8 &   4.4  \\ 
\hline
C   (12) & 168.0 & 161.2 &   5.1  \\
\hline
Si  (28) & 166.3 & 161.8 &   6.7  \\
\hline
Ar  (40) & 165.8 & 162.0 &   7.6  \\
\hline
Sc  (45) & 165.6 & 162.1 &   7.9  \\
\hline
Xe (131) & 164.5 & 162.5 &  11.3  \\
\hline
La (139) & 164.5 & 162.5 &  11.5  \\
\hline
Au (197) & 164.2 & 162.6 &  12.9  \\
\hline
Pb (208) & 164.2 & 162.6 &  13.1  \\
\hline
\end{tabular}
\caption{Estimated limits for the FSS region according to the Ising-QCD description in eq.~(\ref{eq:1}).}
\end{table}

Thus, $T_+$ and $T_-$ define a temperature zone where the critical scaling of baryon number density-density correlation in configuration space is transferred to an analogous scaling for small transverse momenta differences and consequently an intermittency effect in proton transverse momenta is expected to occur. To observe such a behaviour in a system of colliding ions, the temperature of the corresponding freeze-out state has to lie between these two limiting temperatures. This conclusion complements the analysis of \cite{Antoniou2017}. There, it has been shown the narrowness of the critical region in the chemical potential direction ($\Delta \mu_b \approx 5~\mathrm{MeV}$). Here, we show that the FSS region is narrow also along the temperature direction and in fact even narrower for large systems.

\section{Ising-QCD phenomenology in the FSS region}

In this section we will develop a systematic phenomenological study of the FSS region related to the QCD critical point based on two basic ingredients: (i) accurate measurements of the freeze-out parameters for a set of A+A systems at specific collision energy, and (ii) an accurate measurement of $\phi_2$ for an A+A system at the same collision energy, freezing out within the FSS region. Actually both requirements are partially fulfilled by existing experimental data of the NA49 experiment (CERN-SPS) at collision energy $\sqrt{s}=17.2$ GeV. Within these limitations, scaling behaviour compatible with the presence of critical fluctuations is observed in the factorial moments of proton transverse momenta of the Si + "Si"-system (central collisions) \cite{Anticic2015} while, for the same beam energy, the freeze-out parameters for the systems C + C, Si + "Si" and Pb + Pb have been determined in \cite{Becattini2006}. However, these measurements, and in particular the $\phi_2$ measurement in Si + "Si", possess significant statistical errors. To proceed, we will first employ the results of the measurements described previously, ignoring the experimental error of $\phi_2$. Exploiting these (experimental) results we will be able to derive a series of predictions for measurements in the running ion collision experiments at CERN (NA61/Shine) and BNL (RHIC-BES I). We will also discuss the impact of the error in $\phi_2$ on some of these predictions. 

\subsection{Freeze-out parameters}

The chemical potential $\mu_b$ and the temperature $T$ of the freeze-out states generated by central C + C, Si + "Si" and Pb + Pb collisions at $\sqrt{s}=17.2$ GeV are given in reference \cite{Becattini2006}. For completeness we show the corresponding values in Table II, including also the result for the intermittency index $\phi_2$ measurement in the transverse momentum space of protons, produced in central Si + "Si" collisions at the same energy (the errors in $(\mu_b,T)$ are given in the parentheses):
\begin{table}
\begin{tabular}{| l | c | c | c |}
\hline 
System (A) & $T$ (MeV) & $\mu$ (MeV) & $\phi_2$ \\
\hline
\hline
C   (12) & 166.0(4.4) & 262.6(12.9) & - \\
Si  (28) & 162.2(7.9) & 260.0(17.9) & 0.96 \\
Pb (208) & 157.5(2.5) & 248.9(09.0) & - \\
\hline
\end{tabular}
\caption{A+A freeze-out parameters and $\phi_2$, centrality 0-12.5\% (error of $\phi_2$ is ignored)} 
\end{table}  

One can use the central values of Table II to obtain smooth interpolating curves for $T=T(A)$ and $\mu=\mu(A)$ (freeze out parameters as a function of mass number) at the collision energy $\sqrt{s}=17.2$ GeV. It turns out that the function $T=T(A)$ is well described by the relation:
\begin{equation}
T=T_0 + T_a A^{-w_T}~~~~;~~~~T_0=154.56~\mathrm{MeV},~~T_a=37.39~\mathrm{MeV},~~w_T=0.48
\label{eq:8}
\end{equation}
while for the function $\mu=\mu(A)$ we obtain:
\begin{equation}
\mu=\mu_0 + \mu_a e^{-w_{\mu} A}~~~~;~~~~\mu_0=247.36~\mathrm{MeV},~~\mu_a=17.54~\mathrm{MeV},~~w_{\mu}=0.01
\label{eq:9}
\end{equation}
Taking into account also the corresponding errors we obtain an upper and lower envelop in each case.

The expressions (\ref{eq:8},\ref{eq:9}) can now be used to estimate the freeze-out parameters for the Be + Be, Ar + Sc and Xe + La central collisions at $\sqrt{s}=17.2$ GeV in the NA61/SHINE experiment. We find the results given in Table III:
\begin{table}
\begin{tabular}{| l | c | c |}
\hline 
System (A) & $T$ (MeV) & $\mu$ (MeV) \\
\hline
\hline
Be + Be  & 168.5(5.3) & 263.3(12.5)  \\
Ar + Sc  & 160.9(3.3) & 258.0(14.5) \\
Xe + La  & 158.2(2.7) & 251.0(12.4)  \\
\hline
\end{tabular}
\caption{Prediction for NA61/SHINE freeze-out parameters ($\sqrt{s}=17.2~GeV$, central collisions)}
\end{table} 
To obtain the $(T, \mu_b)$-values in Table III for the slightly asymmetric systems Ar + Sc and Xe + La we have used as mass number $A$ the mean between the mass numbers of the two colliding nuclei. The values for the various freeze-out parameters are displayed graphically in Fig.~1(a,b).
In Fig.~1a  the blue circles represent the freeze-out temperature for the central collisions of A+A systems in NA49 experiment (A=C, Si, Pb) while with red circles are shown the predicted freeze out temperatures for A+A' collisions in NA61/SHINE (A=Be, Sc, Xe). The central solid black line is the fit with eq.~(\ref{eq:8}) while the two dashed lines are the lower and upper envelope obtained when taking into account the experimental errors. Similarly in Fig.~1b the blue stars refer to the freeze-out chemical potential of the NA49 systems while the red stars are predictions of the freeze-out chemical potential for the NA61/SHINE systems. The solid black line is the fit with eq.~(\ref{eq:9}) while the dashed lines are upper and lower envelopes induced by the experimental errors. 

\begin{figure}[tbp]
\centering
\includegraphics[width=0.55\textwidth]{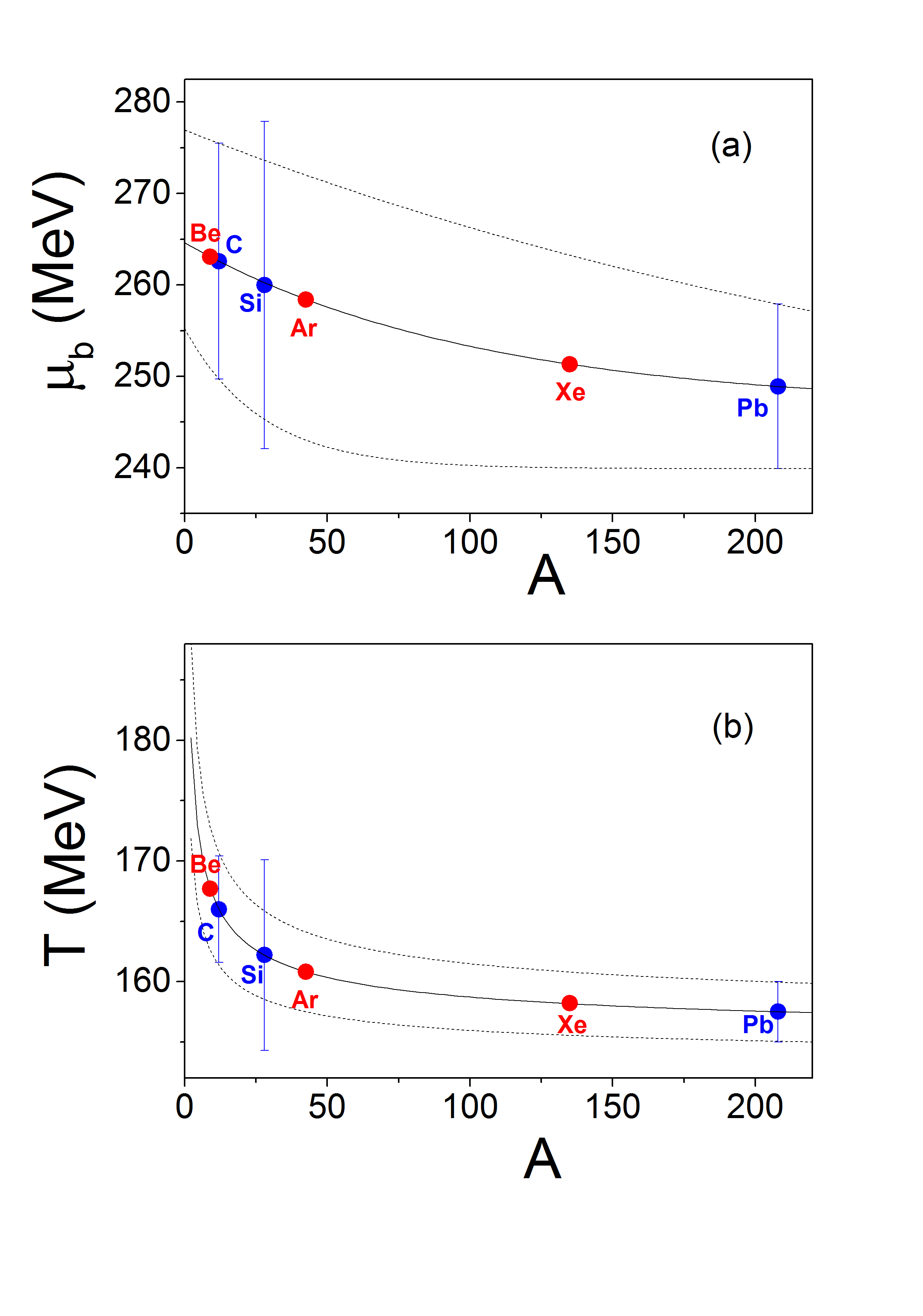}
\caption{(a) The temperature of the freeze-out states for central A+A collisions in NA49 experiment at $\sqrt{s}=17.2$ GeV according to \cite{Becattini2006} are given by the blue circles. The red circles are predictions for the freeze-out temperature of central A+A' collisions in NA61/SHINE experiment ($\sqrt{s}=17.2$ GeV) estimated using eq.~(\ref{eq:8}) (black solid line). The dashed lines are determined by the experimental errors.
(b) Similar to (a) but now for the freeze-out chemical potential. The existing data for NA49 A+A collisions \cite{Becattini2006} are shown by the blue stars while the red stars are predictions for the A+A' collisions in NA61/SHINE experiment. The black solid line represents the graph of eq.~(\ref{eq:9}). Similarly to (a) the two dashed lines are obtained after taking into account the experimental errors.}
\label{fig:fig1ab}
\end{figure} 

Finally, reversing the way of thinking, eq.~(\ref{eq:8}) can be used to determine the range of values for the mass number $A$ for which the condition $T_- \leq T \leq T_c$ is valid. This is a crucial information for ion collision experiments: it provides us with the relevant sizes of nuclei to form freeze-out states within the FSS region, if the colliding energy is $\sqrt{s}=17.2~\mathrm{GeV}$ and the collisions are central. It is straightforward to obtain: $23 \leq A \leq 31$.

\subsection{The intermittency index $\phi_2$}

As a next step we use the NA49 measurement of the intermittency index $\phi_2$ to locate the freeze-out state of the Si + "Si" (central collisions at $\sqrt{s}=17.2$ GeV) within the critical region of the QCD CEP. To achieve this we follow the line of thinking introduced in \cite{Antoniou2017} and summarized briefly in the previous section. There are two conditions for a freeze-out state of A+A ion collisions to lie within the critical region:

\begin{enumerate}
\item The freeze-out temperature $T$ should obey $T_+ < T < T_-$ for the given system's size and
\item the FSS exponent $\tilde{q}$ (and therefore also the intermittency index $\phi_2$) characterizing the scaling of the mean baryon number $\langle N \rangle\sim M^{\tilde{q}}$ with the system's size $M$ should fulfil the condition: $\frac{3}{4} < \tilde{q} < 1$. 
\end{enumerate}

If we ignore measurement uncertainties, both requirements are satisfied by the freeze-out state of central Si + "Si" collisions at $\sqrt{s}=17.2$ GeV since $T=162.2$ MeV \cite{Becattini2006} (while for Si $T_-\approx 161.8$ MeV, $T_+ \approx 166.3$ MeV, see table I) and $\phi_2=0.96$ \cite{Anticic2015}, fulfilling $\frac{3}{4} < \phi_2 < 1$.

We can now employ the partition function (\ref{eq:1}) to determine all the pairs $(\ln \zeta=\frac{\mu_b-\mu_c}{k_B T_c}, t=\vert \frac{T-T_c}{T_c}\vert)$ within the FSS region fulfilling the condition $\tilde{q}=0.96$. To this end we calculate $\langle N \rangle$ for a dense lattice in the $(\ln \zeta,t)$-plane and we determine the points of the lattice for which $\langle N \rangle \sim M^{0.96}$ holds. In fact we find all the $(\ln \zeta, t)$ pairs within the critical region which lead to $\langle N \rangle \sim M^{\tilde{q}}$ with $0.955 < \tilde{q} < 0.965$. The results of this calculation are presented in Fig.~2.    

\begin{figure}[tbp]
\centering
\includegraphics[width=0.55\textwidth]{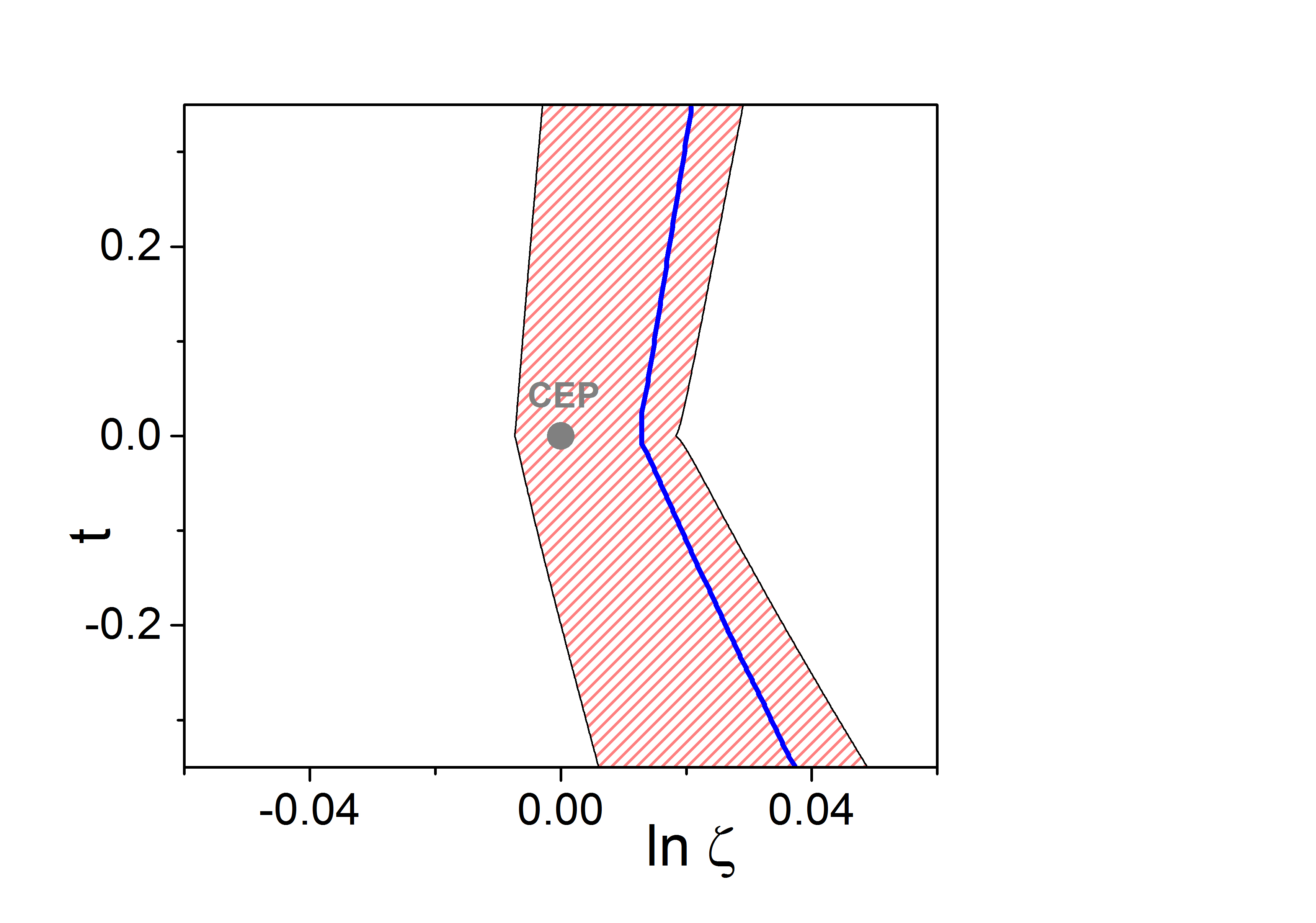}
\caption{The critical region (red shaded area) in the $(\ln \zeta, t)$ plane determined using the partition function in eq.~(\ref{eq:1}). The blue line is the line for which $\tilde{q} \approx 0.96$. The freeze-out state for Si + "Si" central collisions at $\sqrt{s}=17.2~\mathrm{GeV}$ should lie on this blue line.}
\label{fig:fig2}
\end{figure} 

The blue line in Fig.~2 is the line for which $\tilde{q} \approx 0.96$ (with an accuracy of $0.005$). The Si + "Si" freeze-out state in the NA49 experiment (central collisions at $\sqrt{s}=17.2$ GeV) should lie on this blue line which is very well approximated by a piecewise linear function of the form:
\begin{equation}
t^{(Si)} = a^{(Si)}_{\pm} + b^{(Si)}_{\pm} \ln \zeta^{(Si)}
\label{eq:10}
\end{equation} 
with $a^{(Si)}_+ \approx -0.51$, $b^{(Si)}_+ \approx 41.19$ (valid for $t > 0$) and $a^{(Si)}_- \approx 0.17$, $b^{(Si)}_- \approx -14.09$ (valid for $t < 0$). Equation (\ref{eq:10}) introduces a linear relation between $\mu_c$ and $T_c$ which depends on the sign of $t^{(Si)}$:
\begin{equation} 
\mu_c=\left\lbrace \begin{array}{cc} 256.06~MeV + 0.01 T_c &,~~ T^{(Si)} > T_c \\ 271.51~MeV - 0.08 T_c &,~~ T^{(Si)} < T_c \end{array} \right.
\label{eq:11}
\end{equation}
with $T^{(Si)}$ the temperature of the Si + "Si" freeze-out state.
Thus, knowing $T_c$ it allows for the determination of $\mu_c$ and finally opens up the possibility to locate freeze-out states in Fig.~2. We will use the same $T_c$ value as that in Table I ($T_c=163~\mathrm{MeV}$) to calculate the critical chemical potential $\mu_c$ employing eq.~(\ref{eq:11}) and achieve the placement of the freeze-out states of running ion collision experiments on the plane presented in Fig.~2. 
Since $t^{(Si)} < 0$, we have to use the lower branch in eq.~(\ref{eq:11}) obtaining $\mu_c = 257.9$ MeV. We observe that this value is very close to the estimated central value of the freeze-out chemical potential in the Ar + Sc system shown in Table III. The corresponding central value for the freeze-out temperature ($T=160.9~\mathrm{fm}$) lies outside the FSS region. However, the experimental errors for the freeze-out parameters are large enough to allow the entrance into the FSS region as a possibility for Ar + Sc. More accurate measurements of the freeze-out parameters are clearly necessary. This holds for all freeze-out states presented in Tables II and III. On the basis of the central values for $(\mu_b,T)$ in Ar + Sc freeze-out state,  the appearance of an intermittency effect in this system is not expected. On the other hand the freeze-out temperature of Ar + Sc is very close to the corresponding $T_- \approx 162~\mathrm{MeV}$. Therefore, assuming a smooth change of the scaling properties in the freeze-out states departing slightly from the FSS region we expect that the intermittency effect, valid within the FSS region, becomes gradually distorted in a twofold way: 
\begin{itemize}
\item It will occur only up to transverse momentum cells of linear size $\delta p_{\perp}$ fulfilling $\delta p_{\perp} \approx \frac{1}{\delta x_{\perp}}$ with $\delta x_{\perp}$ a scale set by the correlation length of the finite system. The later is expected to be a small fraction of its average transverse size $R_{\perp}$. Therefore, the intermittency effect will gradually disappear with increasing distance from the FSS region. 
\item The value of the associated intermittency index may be altered by boundary effects. 
\end{itemize}
This is in contrast to the behaviour expected to occur for a system freezing out within the FSS region. In this case the intermittency effect should appear as a power-law behaviour of the corresponding factorial moments valid up to transverse momentum cells of linear size $\delta p_{\perp} \approx \frac{1}{R_{\perp}}$ with $R_{\perp}$ the average transverse size of the fireball. Furthermore, the intermittency index $\phi_2$ should be equal to the FSS exponent $\tilde{q}$. Such a behaviour is indeed observed in the intermittency analysis of the Si + "Si" system \cite{Anticic2015} where the power-law behaviour is clearly valid up to transverse momentum cells of linear size $\delta p_{\perp} \approx 30$ MeV, corresponding to a length scale of $6.6~\mathrm{fm}$  which is very close to $R_{\perp}$ for Si (see Table I). Theoretically, for transverse momentum cells with size less than $\frac{1}{R_{\perp}}$ we expect a saturation of the factorial moments to a constant value. This is also compatible with the behaviour seen in Si + "Si", since in this region of $M_{\perp}$ values the factorial moment, after the background subtraction, varies very little and the associated errors are very large  \cite{Anticic2015}. 

Having determined $(\mu_c,T_c)$ it is straightforward to calculate $(\ln \zeta^{(A)}, t^{(A)})$ for the freeze-out states of all the systems presented in Tables II and III. The resulting plot is presented in Fig.~3 and errors in both chemical potential as well as temperature direction are included. In the following discussion we will focus on the central values of the freeze-out parameters which are presented by the colored stars in the plot. In Fig.~3 we have also included the freeze-out state of central Au+Au collisions with $\sqrt{s}=14.5$ GeV generated in RHIC. For the corresponding freeze-out parameters we use the analytic relations given in \cite{Andronic2010}. We observe that the central value lies outside the critical region. Notice that no other freeze-out states from the RHIC BES-I program can fit in this plot. The reason is that the critical region is very narrow in the chemical potential direction, as explained in \cite{Antoniou2017}. On the other hand the beam energy scan at RHIC uses energy steps, which lead to chemical potential values differences which appear very large, compared to the size of the critical region. Coming back to the NA61/SHINE experiment, we observe that, according to Table III, the freeze-out state of the Xe + La system $T_{Xe}=158~\mathrm{MeV}$, $\mu_{Xe}=251~\mathrm{MeV}$ is expected to lie outside the critical region (central value). Similar arguments hold also for the Pb + Pb freeze-out state.

According to this analysis, the critical region can be explored by ion collisions at $\sqrt{s}=17.2~\mathrm{GeV}$ with nuclei with mass number $23 \leq A \leq 31$. Concerning the Ar + Sc system at highest SPS energy we observe that the expected freeze-out state (green star) for central collisions lies within the critical region but outside the FSS region, close to the boundary. Thus, in an intermittency analysis in this system we expect a distortion of the scaling behaviour, along the lines we have described in the previous paragraph. In peripheral Ar + Sc collisions at $\sqrt{s}=17.2~\mathrm{GeV}$ the freeze-out temperature could increase and the chemical potential is expected to remain close to the critical one since the colliding energy does not change. This means that eventually the freeze-out state of peripheral Ar+Sc collisions could enter into the FSS region. In Fig.~3 we omitted for clarity reasons the experimental errors which are very large, especially in the freeze-out chemical potential, indicating the need of more accurate measurements for a safe prediction of the CEP location. Concerning the uncertainty in $\phi_2$ of the Si + "Si", it should be noticed that assuming a value less than $0.96$, which is within the interval dictated by the experimental error, we observe a shift of the location of all freeze-out states to lower values of $\ln \zeta$. In particular, for the limiting value $\phi_2=0.75$, the NA61/SHINE freeze-out states of Ar + Sc and Xe + La collisions at $\sqrt{s}=17.2~\mathrm{GeV}$ would lie outside the critical region. Thus, this relation between the different freeze-out states suggests that accurate $\phi_2$ measurements in NA61/SHINE could also restrict the uncertainty of $\phi_2$ in Si + "Si" system.

\begin{figure}[tbp]
\centering
\includegraphics[width=0.55\textwidth]{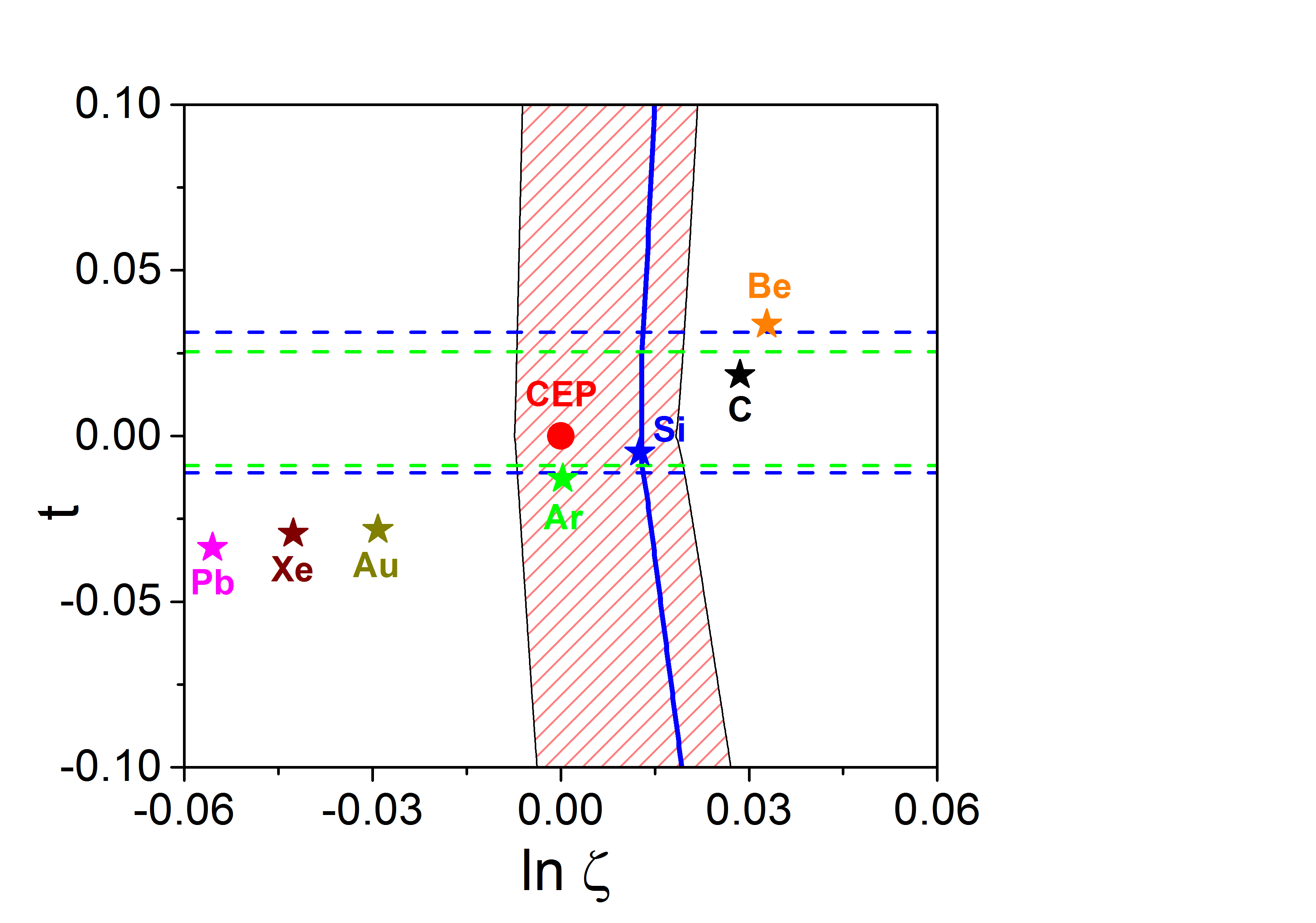}
\caption{The critical region (red shaded area) in the $(\ln \zeta, t)$ plane determined using the partition function in eq.~(\ref{eq:1}) and assuming $T_c \approx 163$ MeV. The colored points indicate the location of the different freeze-out states measured in NA49 experiment (Table II) or expected to be measured in the NA61/SHINE experiment (Table III). The freeze-out state of central Au + Au collisions at $\sqrt{s}=14.5$ GeV is also included (RHIC, BES-I). The blue line is the graphical presentation of $\tilde{q}=0.96$ passing through the freeze-out state of Si + "Si". In general the errors (not shown here), especially for the freeze-out chemical potential,  are very large and  more accurate measurements are needed. The dashed lines (blue, green) present the boarders of the FSS region in each case (Si + "Si", Ar + Sc).}
\label{fig:fig3}
\end{figure} 

\subsection{Global fluctuations, higher order cumulants}

Closing this section it is worth to discuss the possibility to observe critical fluctuations using higher order moments (cumulants) integrated over the entire phase space. Such global measures have been proposed in the literature \cite{Stephanov2009,Athanasiou2010,Stephanov2011} as a tool to observe the CEP in ion collisions. In particular, the non-Gaussian  kurtosis is expected to depend very sensitively on the correlation length when approaching the critical point. As a consequence even a mild maximum of the correlation length will be magnified at the level of kurtosis \cite{Stephanov2009}. Furthermore, it is predicted that kurtosis should be negative close to the critical point \cite{Stephanov2011}. These predictions support the occurrence of a sharp minimum in kurtosis as a function of the ion colliding energy and this strategy in the search for the CEP was adopted by the STAR experiment within the framework of the BES-I program at RHIC \cite{Luo2017}. In \cite{Antoniou2017b} an extensive analysis of the STAR data on kurtosis times variance $\kappa \sigma^2$, along the lines dictated by the Ising-QCD partition function in eq.~(\ref{eq:1}), has been performed. It is argued that the occurrence of minima in kurtosis times variance of baryon number is not a unique signature of the critical point. Furthermore, in \cite{Antoniou2017b} it has been argued that the quantity $\kappa \sigma^2$ is a good candidate for the description of fluctuations outside the critical region, while within the critical region the relevant quantity is the kurtosis itself. In fact, this can be shown rigorously, employing a general grand-canonical description of the partition function of the baryon system far from the critical point:
\begin{equation}
\mathcal{Z}=\displaystyle{\sum_{N}} e^{N \ln \zeta} \exp[-F(N,V,T)]
\label{eq:12}
\end{equation}
with $\zeta=\frac{\mu_b - \mu_c}{T_c}$ and $V$ the system's volume as in eq.~(\ref{eq:1}). 
Then, all the higher cumulants $C_k$ of the baryon number $N$ are expressed as:
\begin{equation}
C_k = \frac{\partial^k}{\partial (\ln \zeta)^k} \ln \mathcal{Z}~~~~;~~~~k=2,~3,..
\label{eq:13}
\end{equation}
as it can be proven easily by induction. Therefore, since $\ln \mathcal{Z}$ (Gibbs free energy) is proportional to the volume $V$ of the system and $\ln \zeta$ is an intensive quantity, the cumulants $C_k$ are proportional to $V$. Consequently, non-Gaussian kurtosis $\kappa=\frac{C_4}{C_2^2}$ is proportional to $\frac{1}{V}$ and $\kappa \sigma^2$ (with $\sigma^2=C_2$) is a size independent quantity. This description breaks down in the critical region due to the anomalous scaling in eq.~(\ref{eq:3}). There, $\kappa$ itself is size independent and it should be used as a measure of fluctuations for freeze-out states within the FSS region. In fact, the baryon number kurtosis, within the Ising-QCD description, has been calculated in \cite{Antoniou2017b} and it has been shown that it possesses a very sharp minimum, becoming at the same time negative, in accordance with the theoretical expectations discussed above. However, the width of this minimum in the beam energy direction is extremely tiny. This is related to the narrowness of the critical region along the chemical potential direction. Thus, since the BES-I program at RHIC explores the QCD phase diagram in the ($\mu_b$,$T$) plane with minimal step-size $\Delta \mu_b \approx 50~\mathrm{MeV}$ it is very unlikely to detect  fluctuations characterizing the critical region which has a size of $\approx 5~\mathrm{MeV}$ along the baryochemical potential axis.

\section{Concluding remarks}

We have performed a detailed phenomenological study of the critical region around the QCD critical endpoint using two basic tools:  measurements of the freeze-out parameters for central ion collisions at $\sqrt{s}=17.2$ GeV and measurements of the corresponding intermittency index $\phi_2$ using proton transverse momenta at midrapidity. Our analysis, based on the connection of intermittency in transverse momentum space with the FSS in configuration space, as revealed in \cite{Antoniou2017}, leads to an estimate of the critical chemical potential value $\mu_c \approx 258$ MeV and a series of predictions for expected experimental results in eventual measurements of the intermittency index $\phi_2$ in running experiments.

Furthermore, we find that the freeze-out chemical potential of the Ar + Sc system (central collisions at $\sqrt{s}=17.2$ GeV) is expected to lie very close to the critical chemical potential. However, the corresponding freeze-out temperature will lie outside the FSS region suppressing the intermittency phenomenon in transverse momentum space. This is due to the fact that the FSS region is very narrow along the temperature axis. This is a crucial observation complementary to the one made in \cite{Antoniou2017}: there, it has been shown that the critical region is very narrow along the baryochemical direction. Here we have shown that the finite-size scaling region, subdomain of the critical region, is very narrow also along the temperature direction. This makes the $\phi_2$-measurement in Si + "Si" collisions at the NA49 experiment of exceptional importance, since this process appears to freeze-out within the FSS region. Very accurate measurements of $\phi_2$ in central Si + "Si" collisions at $\sqrt{s}=17.2~\mathrm{GeV}$ could provide us with invaluable information concerning the location of the CEP. 

Furthermore, we emphasize that the scheduled energy variation in the BES-I program at RHIC, leading to steps of at least $50~\mathrm{MeV}$ along the chemical potential axis, cannot capture the conditions necessary for the entrance into the critical region with a size of $\approx 5~\mathrm{MeV}$ in $\mu_b$ direction \cite{Antoniou2017b}.  Thus, to produce freeze-out states within the critical region, a high resolution for the freeze-out chemical potential is required, which can only be achieved using colliding nuclei of varying size. 

According to our analysis, potentially promising measurements for observing critical fluctuations are linked to peripheral Ar + Sc collisions in the NA61/SHINE experiment at maximum SPS energy. Accurate measurements of the freeze-out state parameters and the intermittency index $\phi_2$ in these collisions, as well as in systems with mass number $23 \leq A \leq 31$, are of high priority.

\end{document}